\documentclass[epjCONF]{svjour}
\usepackage{amsmath}
\usepackage{graphics}
\usepackage[varg]{txfonts} 
\usepackage[latin1]{inputenc}
\usepackage[final]{graphicx}

\graphicspath{{./pics/}}
\DeclareGraphicsExtensions{.eps}

\session-title{Hot and Cold Baryonic Matter -- HCBM 2010}
\begin{document}
\title{Collective Flow and Mach Cones with parton transport}
\author{I.~Bouras\inst{1} \and A.~El\inst{1} \and O.~Fochler\inst{1} \and F.~Reining\inst{1}
\and J.~Uphoff\inst{1} \and C.~Wesp\inst{1} \and Z.~Xu\inst{1,2} \and
C.~Greiner\inst{1}\fnmsep\thanks{\email{carsten.greiner@th.physik.uni-frankfurt.de}}
}
\institute{
Institut fuer Theoretische Physik,
Goethe Universitaet Frankfurt, Max-von-Laue-Strasse.1,
D-60438 Frankfurt am Main, Germany
\and
Frankfurt Institute for Advanced Studies,
Ruth-Moufang-Strasse 1,
D-60438 Frankfurt am Main, Germany
}
\abstract{
Fast thermalization and a strong build up of elliptic flow
of QCD matter were investigated within the pQCD based
3+1 dimensional parton transport model BAMPS including
bremsstrahlung $2 \leftrightarrow 3$ processes.
Within the same framework quenching of gluonic jets
in Au+Au collisions at RHIC can be understood.
The development of conical structure by gluonic jets is
investigated in a static box for the regimes of small and
large dissipation. Furthermore we demonstrate two different
approaches to extract the shear viscosity coefficient $\eta$
from a microscopical picture.
} 
\maketitle
%

\section{Introduction}

The large value of the elliptic flow $v_2$ measured in experiments at the Relativisitc
Heavy Ion collider (RHIC)\cite{Adler:2003kt,Adams:2003am,Back:2004mh} suggests, that
in the hot and dense fireball equilibration of quarks and gluons occurs on a very short
time scale $\le 1$ fm/c. This results into the assumption,
that this locally thermalized matter, the quark gluon plasma (QGP), behaves
like a nearly perfect fluid. In this case quarks and gluons should be rather
strongly coupled and the QGP should have a very small viscosity to entropy ratio $\eta/s$.
This value might in fact be close to the conjectured lower bound $\eta/s = 1/(4 \pi)$, obtained from a
correspondence between conformal field theory (CFT) and string theory in an
Anti-de-Sitter space (AdS) \cite{Kovtun:2004de}.

The phenomenon of jet-quenching was another important discovery
at RHIC \cite{Adams:2003kv}. Hadrons with high transverse momenta are suppressed in
$Au + Au$ collisions with respect to a scaled $p + p$ reference
\cite{Adler:2002xw,Adcox:2001jp}. This quenching of jets is commonly
attributed to energy loss on the partonic level as the hard partons
produced in initial interactions are bound to traverse the QGP created in the 
early stages of heavy-ion collisions (HIC). In addition, very exciting jet-associated particle
correlations were observed \cite{Wang:2004kfa}, which might be the result of a
conical emission off propagating shock waves in form of Mach Cones. These Mach Cones might be
induced by high-energetic partons traversing the expanding medium \cite{Stoecker:2004qu}. 

A large class of phenomena in heavy-ion collisions can be investigated
within the framework of the kinetic transport theory. Among others,
the relativistic pQCD-based on-shell
parton transport model BAMPS (Boltzmann Approach to Multiparton Scatterings)
\cite{Xu:2004mz} was developed to describe the
early QGP stage of a heavy-ion collision (HIC). Using BAMPS early thermalization
of gluons within $\tau < 1$ fm/c was demonstrated in Au+Au collisions at
$\sqrt{s_{NN}} = 200$ GeV
employing Glauber initial conditions of minijets and the coupling constant
$\alpha_s = 0.3 - 0.6$. In addition to the
elastic pQCD $gg \leftrightarrow gg$ processes, pQCD-based
bremsstrahlung $gg \leftrightarrow ggg$ was included. These were shown
to be essential for the achievement of local thermal equilibrium
at that short time scale. The fast thermalization happens also in a
similar way with Color Glass Condensate initial conditions \cite{El:2007vg}.

BAMPS has been recently applied to calculate elliptic flow and
jet quenching at RHIC energies \cite{Fochler:2008ts} for the first time
in a consistent and fully pQCD--based microscopic transport model.
Both key observables could be addressed on the
partonic level within a common setup.
The left panel of Fig. \ref{fig:v2_summary_RAA_central} shows that the
medium simulated in the parton cascade BAMPS exhibits a sizable degree
of elliptic flow in agreement with experimental findings at RHIC
as discussed in Ref. \cite{Xu:2007jv,Xu:2008av}.

Extraction of the shear viscosity over entropy density ratio $\eta/s$ from BAMPS simulations
confirmed the essential importance of inelastic processes. Bremsstrahlung and back reaction
processes lower the shear viscosity to entropy density ratio significantly, by a
factor of $7$, compared to the values obtained if only elastic collisions are
considered \cite{Xu:2007ns,El:2008yy}. For
$\alpha_s = 0.3$ one finds $\eta/s = 0.13$, whereas for $\alpha_S = 0.6$ the obtained value
matches the lower bound of $\eta/s = 1/(4\pi)$ from the AdS/CFT conjecture.

In these proceedings we show some recent descriptions and developments
of different phenomena in
relativistic HIC using BAMPS. Due to the large momentum scales
involved, the energy loss of partonic jets can be treated in terms of
perturbative QCD (pQCD) and most theoretical schemes attribute the main
contribution to partonic energy loss to radiative processes
\cite{Wicks:2005gt}. In addition, the possible
propagation of Mach Cones in the QGP induced by such high-energetic
partons can be studied. Considering the earlier works, in which
the effects of dissipation on relativistic shock waves were investigated
\cite{Bouras:2009nn,Bouras:2009vs,Bouras:2010hm}, we demonstrate
the transition of Mach Cones from ideal to the viscous ones.
It is a major challenge to combine jet physics on the one hand and
bulk evolution on the other hand within a common framework.
In the end we demonstrate two new independent methods to extract numerically the
shear viscosity to entropy density ratio $\eta/s$. One of them is based
on the classical picture of a linear velocity gradient, where the
shear viscosity coefficient $\eta$ can be calculated from the
local deformation of the distribution function. The other method is
based on the Green-Kubo relation, where shear viscosity is obtained
from the autocorrelation function of equilibrium fluctuations.

\section{Jet Quenching in Au+Au collisions at 200 AGeV}

For simulations of Jet Quenching in heavy ion collisions the initial gluon
distributions are sampled according to a mini--jet model with a lower
momentum cut-off $p_{0} = 1.4$ GeV and a $K$--factor of $2$.
The test particle method \cite{Xu:2004mz} is employed to ensure sufficient
statistics.
Quarks are discarded after sampling the initial parton distribution since currently
a purely gluonic medium is considered. To model the freeze out of the
simulated fireball, free streaming is applied to regions where the local
energy density has dropped below a critical energy density
$\varepsilon_{c}$ ($\varepsilon_{c} = 1.0\, \mathrm{GeV}/\mathrm{fm}^3$
unless otherwise noted).

\begin{figure*}[tbh]
  \centering
  \begin{minipage}[t]{0.42\textwidth}
    \includegraphics[width=\linewidth]{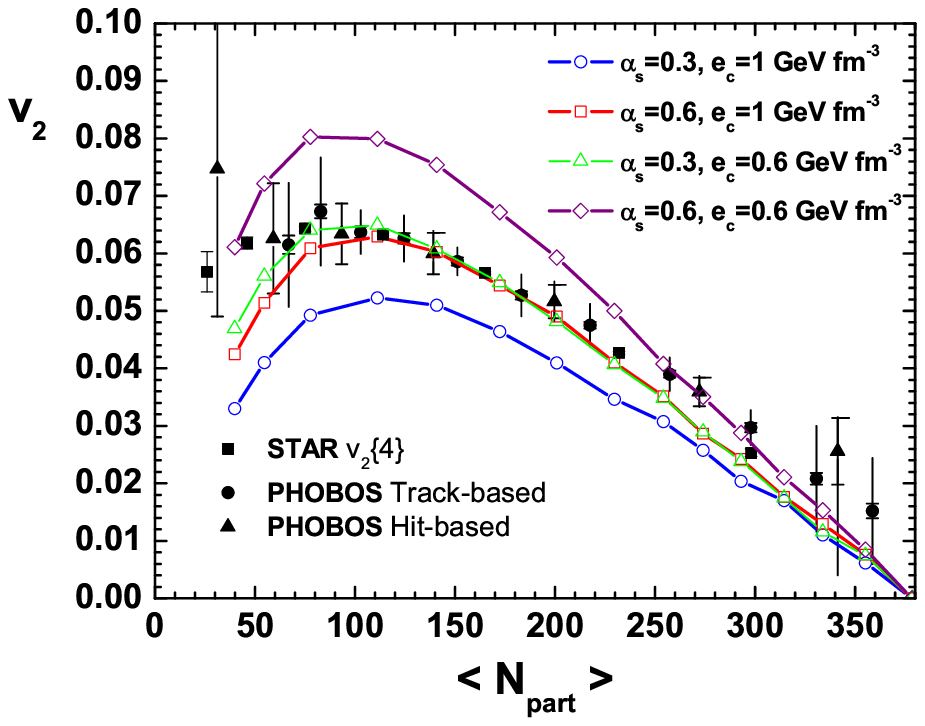}
  \end{minipage}
  \begin{minipage}[t]{0.48\textwidth}
    \includegraphics[width=\linewidth]{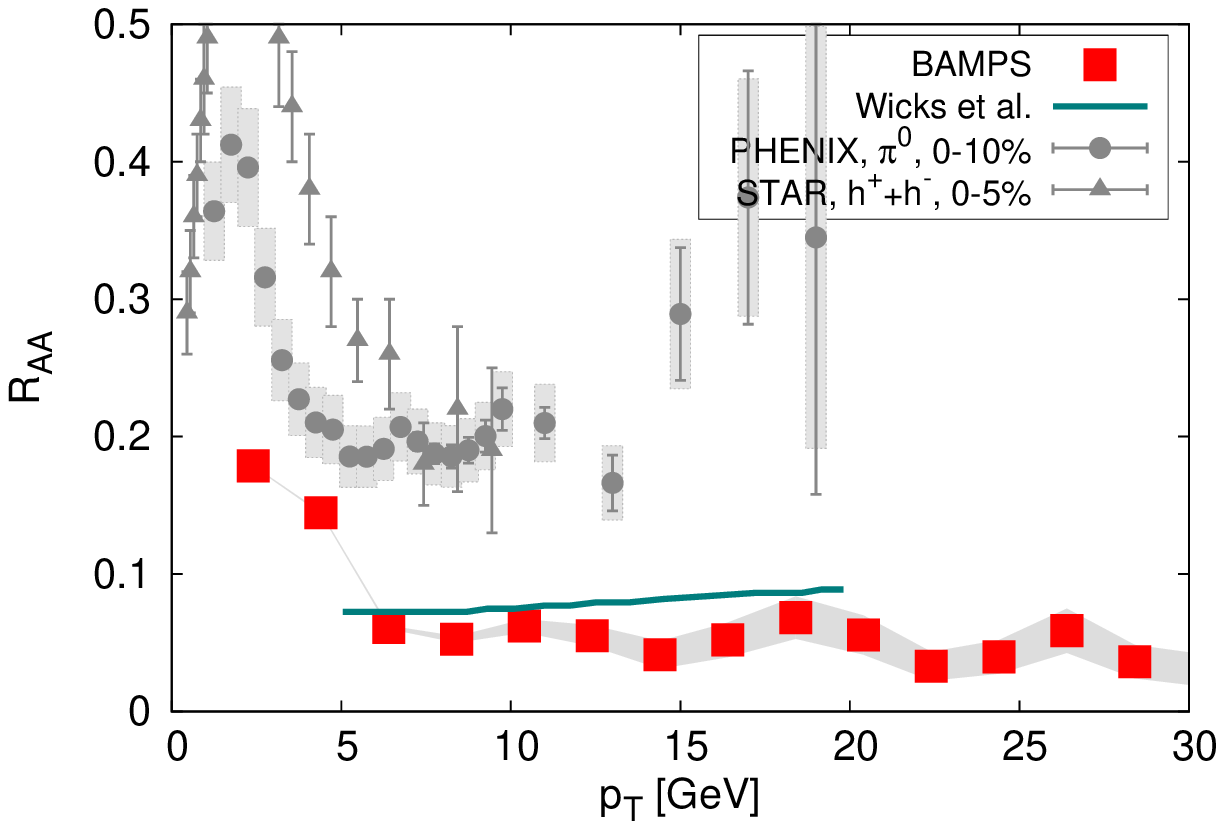}
  \end{minipage}

  \caption{Left panel: Elliptic flow $v_2$ as a function of the number
of participants for Au+Au at 200~AGeV for different combinations of the
strong coupling $\alpha_s$ and the critical energy density $\varepsilon_c$.
See \cite{Xu:2008av} for more information. \newline Right panel: Gluonic
$R_{AA}$ at midrapidity ($y \,\epsilon\, [-0.5,0.5]$) as extracted from
simulations for central Au+Au collisions at 200~AGeV. For comparison the
result from Wicks et al. \cite{Wicks:2005gt} for the gluonic contribution
to $R_{AA}$ and experimental results from PHENIX \cite{Adare:2008qa} for
$\pi^{0}$ and STAR \cite{Adams:2003kv} for charged hadrons are shown.}
\label{fig:v2_summary_RAA_central}
\end{figure*}

The right panel of Fig. \ref{fig:v2_summary_RAA_central} shows the gluonic
$R_{AA}$ simulated in BAMPS for central,  $b=0\,\mathrm{fm}$, collisions.
It is roughly constant at $R_{AA}^{\mathrm{gluons}} \approx 0.053$ and in
reasonable agreement with analytic results for the gluonic 
contribution to the nuclear modification factor $R_{AA}$ \cite{Wicks:2005gt},
though the suppression of gluon jets in BAMPS appears to be slightly stronger.
We expect improved agreement in future studies when employing a carefully
averaged $\langle b \rangle$ that will be better suited for comparison to
experimental data than the strict $b=0\,\mathrm{fm}$ case.

\begin{figure*}[tbh]
  \centering
  \begin{minipage}[l]{0.45\textwidth}
    \includegraphics[angle=270,width=\linewidth]{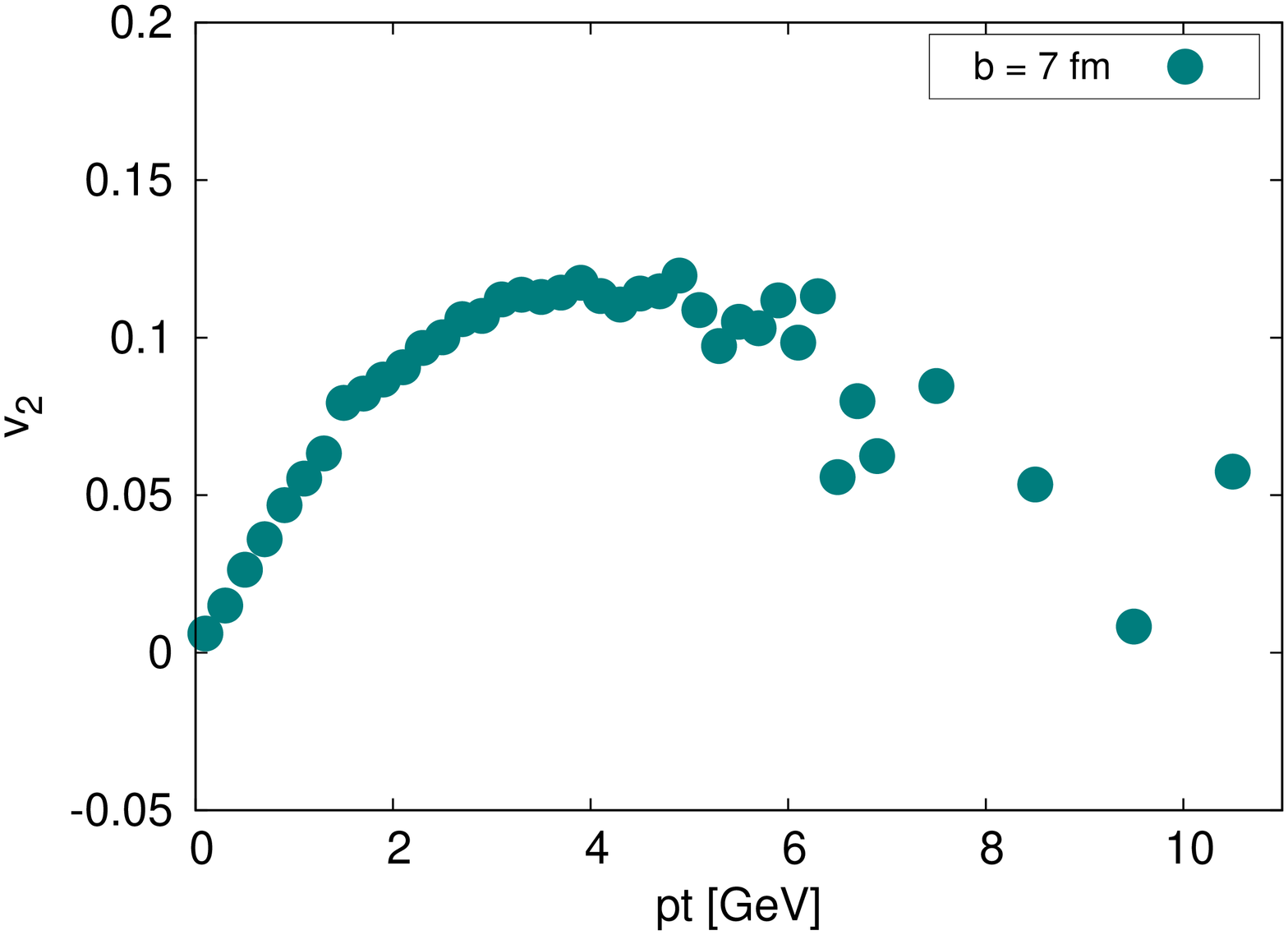}
  \end{minipage}
  \begin{minipage}[r]{0.45\textwidth}
    \includegraphics[width=\linewidth]{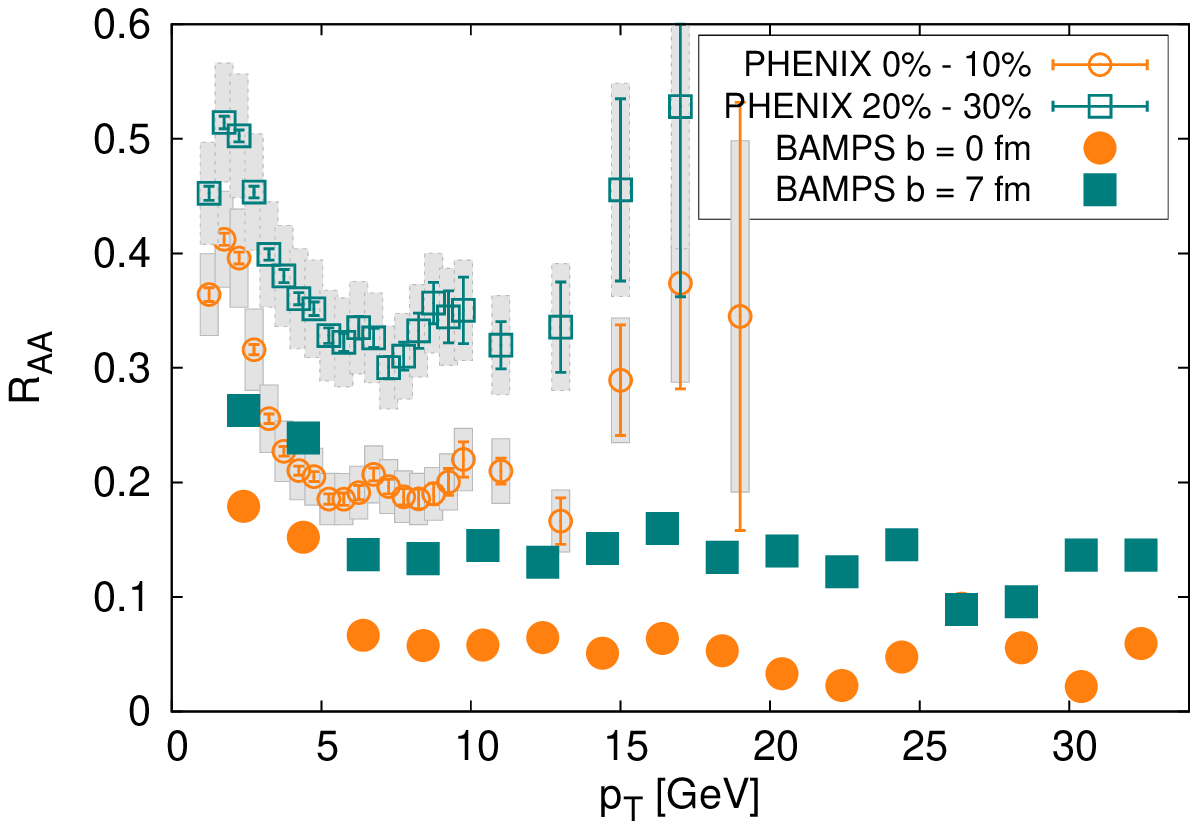}
  \end{minipage}
  \caption{Left panel: Elliptic flow $v_{2}$ for gluons in simulated
Au+Au collisions at 200 AGeV with $b=7\,\mathrm{fm}$. $\varepsilon_c = 0.6\, \mathrm{GeV}/\mathrm{fm}^3$.
See \cite{Fochler:2010wn} fore information. \newline
Right panel: Gluonic $R_{AA}$ as extracted from BAMPS simulations for
$b=0\,\mathrm{fm}$ and $b=7\,\mathrm{fm}$, $\varepsilon_c = 1.0\, \mathrm{GeV}/\mathrm{fm}^3$.
For comparison experimental results from PHENIX \cite{Adare:2008qa} for $\pi^{0}$ are shown for
central $(0 - 10) \%$ and off--central $(20-30)\%$ collisions.}
\label{fig:v2_RAA_b7}
\end{figure*}

We have computed the gluonic $R_{AA}$ for non--central Au + Au collisions
at the RHIC energy of $\sqrt{s} = 200 \mathrm{AGeV}$ with a fixed impact
parameter $b=7\,\mathrm{fm}$ (Fig. \ref{fig:v2_RAA_b7}), which roughly
corresponds to $(20-30)\%$ experimental centrality. A comparison in
terms of the magnitude of the jet suppression for $b=7\,\mathrm{fm}$ is
difficult since there are no published analytic results available to compare
to. Taking the ratio of the $b=7\,\mathrm{fm}$ to the $b=0\,\mathrm{fm}$
results as a rough guess indicates that the decrease in quenching is more
pronounced in BAMPS compared to experimental data. The ratio of the nuclear
modification factor between central $(0 - 10) \%$ and more peripheral
$(20-30)\%$ collisions is $\left. R_{AA}\right|_{0 \% - 10 \%} / \left. R_{AA}\right|_{20 \% - 30 \%} \approx 0.6$
for the experimental data, while for the BAMPS results
$\left. R_{AA}\right|_{b=0\,\mathrm{fm}} / \left. R_{AA}\right|_{b=7\,\mathrm{fm}} \approx 0.4$.
However, the issue of detailed quantitative comparison needs to be
re-investigated once light quarks and a fragmentation scheme are
included into the simulations.

To complement the investigations of $R_{AA}$ at a non--zero impact parameter
$b=7\,\mathrm{GeV}$, we have computed the elliptic flow parameter $v_{2}$
for gluons at the same impact parameter and extended the range in transverse
momentum up to roughly $p_{T} \approx 10\,\mathrm{GeV}$, see left panel of
Fig. \ref{fig:v2_RAA_b7}. For this a critical energy density
$\varepsilon_{c} = 0.6\, \mathrm{GeV}/\mathrm{fm}^3$ has been used, in order
to be comparable to previous calculations of the elliptic flow within BAMPS.
The $v_{2}$ of high--$p_{T}$ gluons is at first rising with $p_{T}$, but from
$p_{T} \approx 4$ to $5\,\mathrm{GeV}$ on, it begins to slightly decreases again.
This behavior is in good qualitative agreement with recent RHIC data
\cite{Abelev:2008ed} that for charged hadrons shows $v_{2}$ to be rising up
to $v_{2} \approx 0.15$ at $p_{T} \approx 3\,\mathrm{GeV}$ followed by a
slight decrease.


\section{Transition from ideal to dissipative Mach Cones}

In the early 1970s shock waves were theoretically predicted
to occur in relativistic heavy-ion collisions (HIC)
\cite{PhysRevLett.32.741} and they have been experimentally
investigated. There is an
important issue in recent studies whether the small but finite
viscosity allows the development of relativistic shocks in form
of Mach Cones in such a hot and dense matter like the QGP.
Within the framework of BAMPS studies were finished
to answer the question whether shocks can develop
with finite viscosity and how this will alter such
a picture \cite{Bouras:2009nn}. Within the relativistic
Riemann problem it was shown that one dimensional shocks
smears out if viscosity is large
\cite{Bouras:2009vs,Bouras:2010hm}. However, the expected
viscosity in HIC seems to be small enough to allow a
significant contribution of shocks in form of Mach Cones.
In the following we report a very recent study.

Mach Cones, which are special phenomena of shock waves,
have their origin in ideal hydrodynamics. A very weak
perturbation in a perfect fluid induces sound waves which
propagate with the speed of sound $c_s = \sqrt{dp/de}$,
where $p$ is the pressure and $e$ is the energy density.
In the case where the perturbation with velocity $v_{\rm jet}$
propagates faster than the generated sound waves, the sound
waves lie on a cone. Considering a gas of massless particles,
where $e = 3p$ and $c_s = 1/\sqrt{3}$, then the emission angle
of the Mach Cone is given by
$\alpha_w = \arccos ( c_s / v_{\rm jet} ) = 54,73^\circ$.

%
\begin{figure*}[ht]
\centering 
\includegraphics[width=\textwidth]{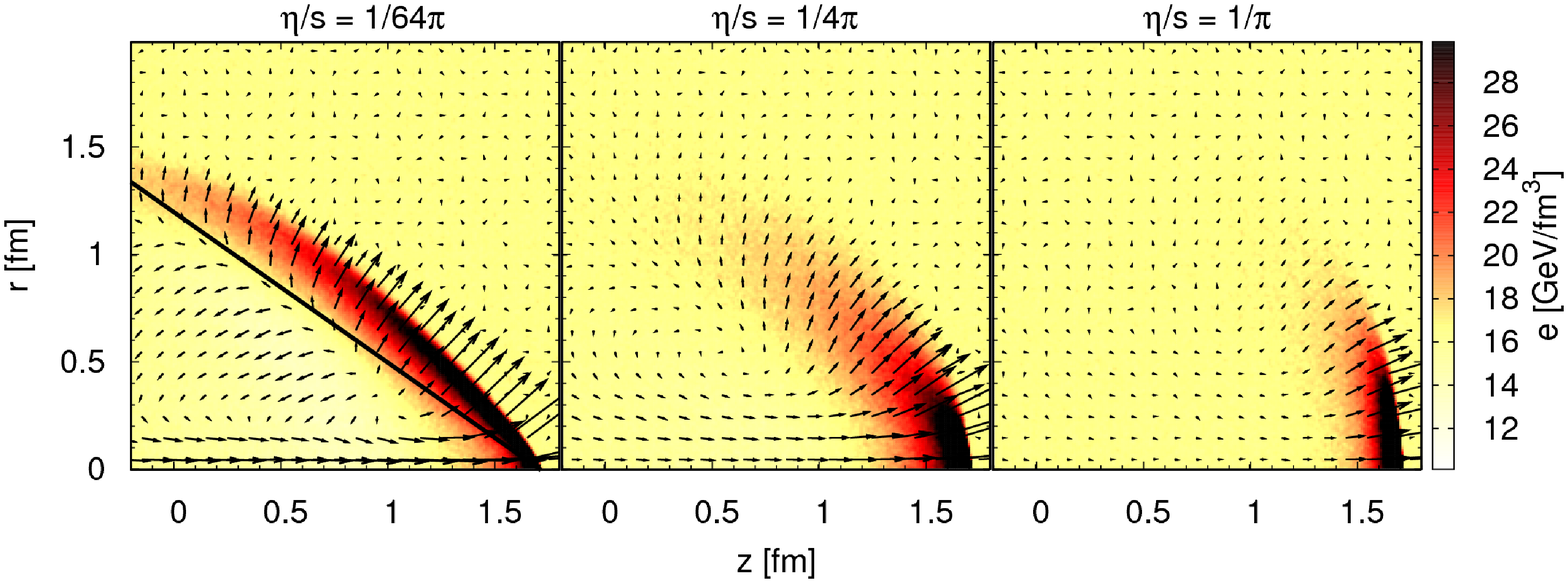}
\caption{(Color online) Scenario of a massless
jet with $p_z = E_{\rm jet} = 200$ GeV in the punch through
scenario - the shape
of a Mach Cone shown for different viscosities of the medium,
$\eta/s = 1/64\pi$ (left), $\eta/s = 1/4\pi$
(middle), $\eta/s = 1/\pi$ (right). The energy deposition of
the jet is approximately $dE/dx = 11 - 14$ GeV/fm. We show
the energy density plotted together with the velocity profile.
Additionally, in the left panel the linear ideal Mach Cone for
a very weak perturbation is shown by a solid line; its emission
angle is $\alpha_w = 54,73^\circ$.}
\label{fig:machCone}
\end{figure*}
%

A stronger perturbation induces the propagation of shock waves
exceeding the speed of sound, therefore the emission angle
changes and can be approximated by\\
$\alpha \approx \arccos ( v_{\rm shock} / v_{\rm jet} )$. Here 
\begin{equation*}
\label{eq:v_shock}
v_{\rm{shock}} = \left [ \frac{(p_1 - p_0)(e_0 + p_1)}
{(e_1 - e_0)(e_1 + p_0)} \right]^{\frac{1}{2}}
\end{equation*}
is the velocity of the shock front, $p_{ \rm 0}$ ($e_{ \rm 0}$) 
the pressure (energy density) in the shock front region and
$p_{ \rm 1}$ ($e_{ \rm 1}$) in the stationary medium itself.
The expression has the following limits: For
$p_0 >> p_1$ we obtain $v_{\rm{shock}} \approx 1$, whereas
for a small perturbation, $p_0 \approx p_1$, we get
$v_{\rm{shock}} \approx c_s$.

We employ the microscopic transport model BAMPS to investigate Mach
Cones with different strength of dissipations in the medium using a
jet moving in positive $z$-direction,
initialized at $t = 0$ fm/c at the position $z = -0.8$ fm.
The jet is treated as a massless particle with zero
spatial volume and zero transverse momentum, that is, $p_z = E_{\rm jet} = 200$
GeV and $v_{\rm jet} = 1$. The energy and momentum deposition to the
medium is realized via collisions with medium particles. In this scenario
we neglect the deflection of the jet and it can not be stopped by the medium;
its energy and momentum is set to its initial value after every collision.

%
\begin{figure*}[th]
\includegraphics[width=\textwidth]{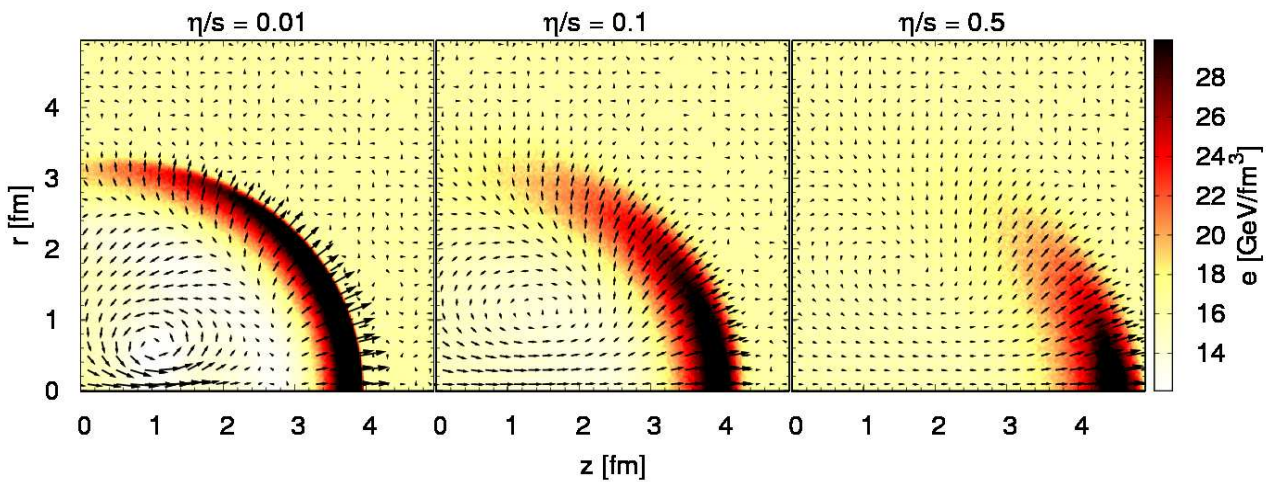}
\caption{(Color online) Scenarion of a deflectable jet with finite energy
$p_z = E_{\rm jet} = 20$ GeV - Induced Mach Cone structure
for different viscosities of the medium, $\eta/s = 0.01$ (left),
$\eta/s = 0.1$ (middle), $\eta/s = 0.5$ (right). We show the
energy density plotted together with the velocity profile.}
\label{fig:machCone_jetBundle}
\end{figure*}
%

All simulations are
realized within a static and uniform medium of massless Boltzmann
particles and $T = 400$ MeV. For this study we consider only binary scattering
processes with an isotropic cross section among the bulk particles.
To save computational
runtime we reduce our problem to two dimensions. Here we choose
the $xz$-plane and apply a periodic boundary condition in $y$-direction. 

In Fig.\ref{fig:machCone} we demonstrate the transition from ideal
Mach Cone to a highly viscous one by adjusting the shear viscosity
over entropy density ratio in the medium from
$\eta/s = 1/64 \pi \approx 0.005$ to $1/\pi \approx 0.32$.
The energy deposition of the jet is approximately $dE/dx = 11 - 14$
GeV/fm. We show a snapshot at $t = 2.5$ fm/c.

Using a small viscosity of $\eta/s = 1/64\pi$, we
observe a strong collective behavior in form of a Mach Cone,
as shown in the left panel of Fig.\ref{fig:machCone}.
Due to the fact that the energy deposition is strong,
the shock propagates faster than the speed of sound through
the medium. For comparison, the ideal Mach Cone caused by a very weak
perturbation is given by a solid line with its emission angle
$\alpha_w = 54,73^\circ$.
Furthermore, a strong diffusion wake in direction of the jet,
characterized by decreased energy density, and a head shock in
the front are clearly visible.

If we increase the viscosity of the medium to larger values, shown
in the middle and left panel of Fig.\ref{fig:machCone}, the
typical Mach Cone structure smears out and vanishes completely.
Due to stronger dissipation, the collective behavior gets weaker
because of less particle interactions in the medium with a larger
$\eta/s$. The results agree qualitatively with earlier studies
\cite{Bouras:2009nn,Bouras:2010hm},
where a smearing-out of the shock profile is observed with
higher viscosity.

In addition to the scenario of a punch through jet we demonstrate in
Fig.\ref{fig:machCone_jetBundle} the scenario of a massless jet
with finite energy which can also be deflected.
Its initial energy is set to $p_z = E_{\rm jet} = 20$ GeV, where the
starting point is $z = -0.3$ fm. We show the results for different
viscosities, $\eta/s \approx 0.01$ to $0.5$ at $t = 5.0$ fm/c.
In analogy to the results above we observe a clear Mach Cone
structure for small viscosities and a smearing out with larger values
of $\eta/s$. Only in the ideal case a strongly curved structure in which the
building up of a strong vortex is visible. The physical meaning
of these phenomena and also jets with the full pQCD cascade
have to be explored in future studies \cite{bourasMachCones}.


\section{Extraction of shear viscosity from microscopic theory}

\begin{figure*}[tbh]
  \centering
  \begin{minipage}[l]{0.45\textwidth}
    \includegraphics[width=\linewidth]{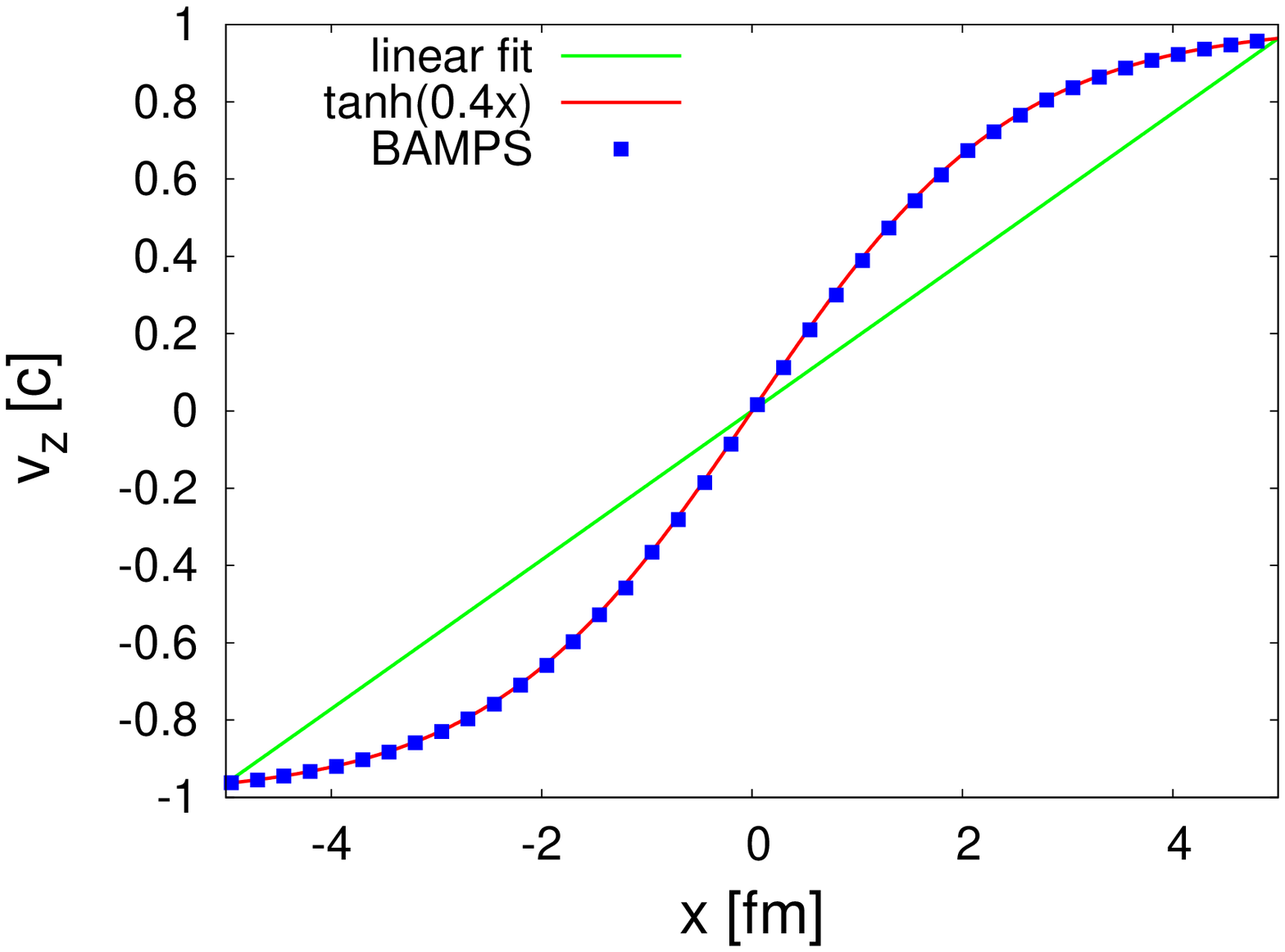}
  \end{minipage}
  \begin{minipage}[r]{0.45\textwidth}
    \includegraphics[width=\linewidth]{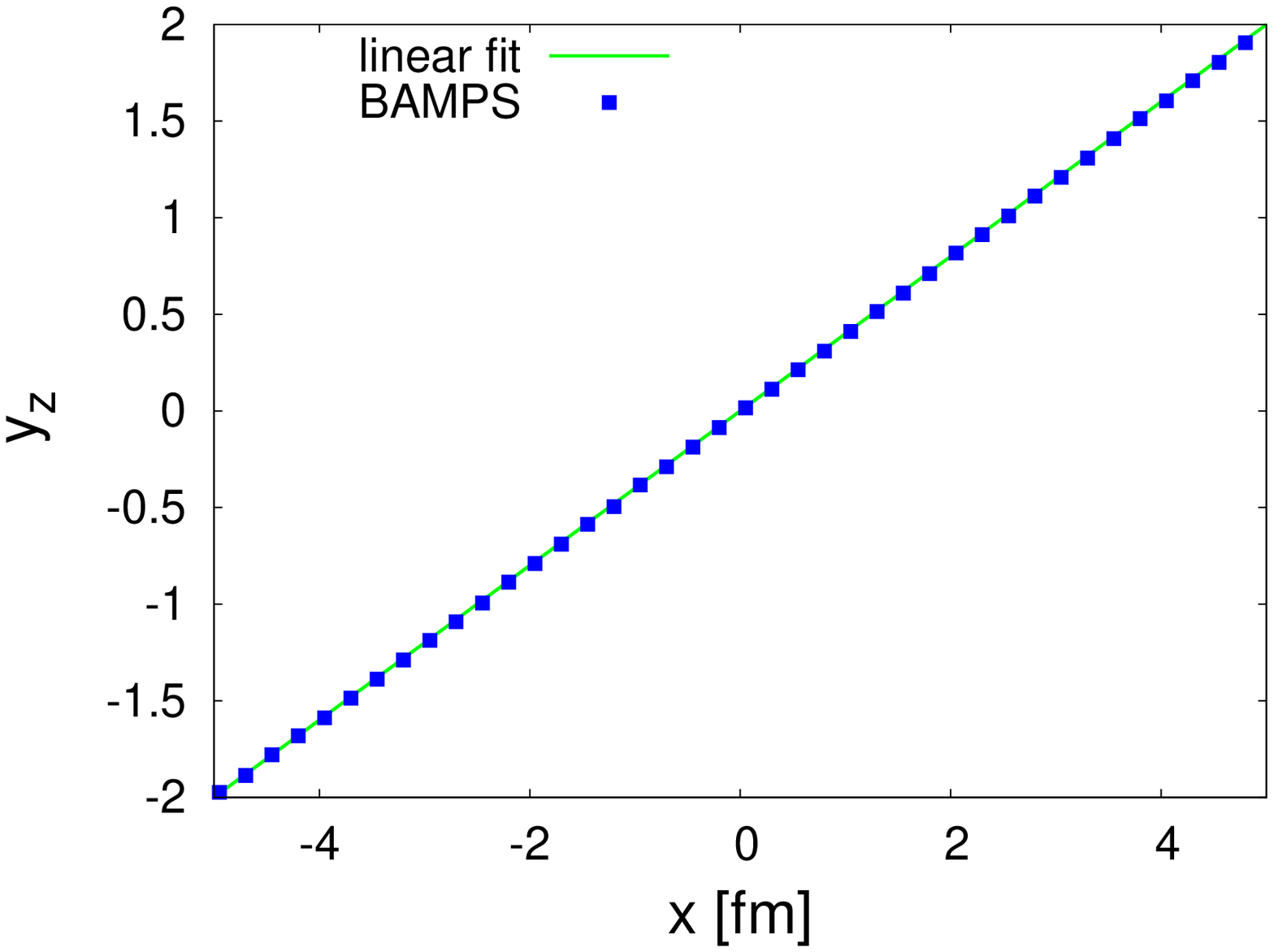}
  \end{minipage}
  \caption{On the left panel we show the velocity profile $v_{\rm z}(x)$
extracted from BAMPS compared to a linear fit and a hyperbolic tangent,
on the right panel we show the according rapidity profile $y_{\rm z}(x)$.
We use $T = 0.4$ GeV, $L=10$ fm, $\lambda_{\rm mfp} = 0.01$ and
$v_{\rm wall} = 0.964$.}
\label{fig:velocityGradient}
\end{figure*}
%

%
\begin{figure*}[th]
  \centering
  \begin{minipage}[l]{0.45\textwidth}
    \includegraphics[width=\linewidth]{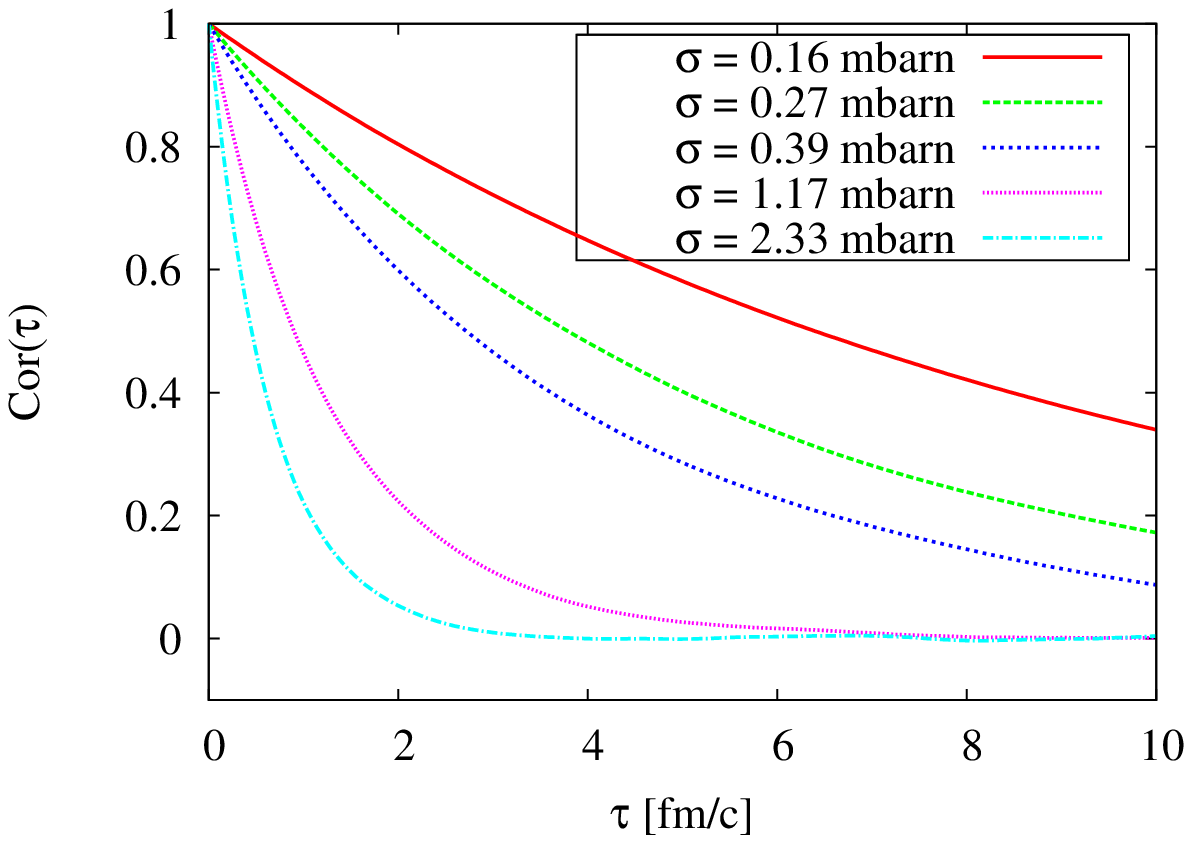}
  \end{minipage}
  \begin{minipage}[r]{0.45\textwidth}
    \includegraphics[width=\linewidth]{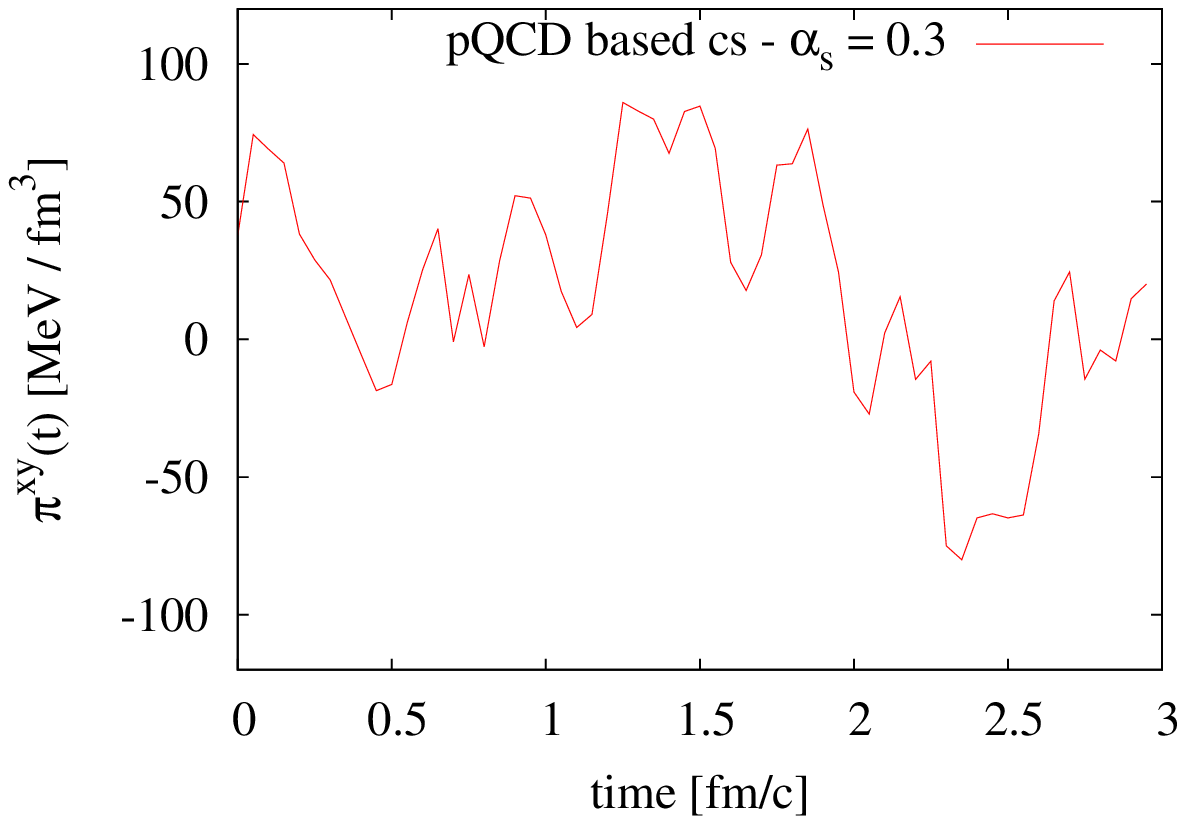}
  \end{minipage}
  \caption{Left panel: The correlation function for different
isotropic cross section is shown. The relaxation time $\tau$
is linear dependent on the cross section $\sigma$.\newline
Right panel: A snapshot of a fluctuating shear stress tensor, which is used to
calculate (in this case for pQCD interactions) the correlation function.}
\label{fig:iso_cs22}
\end{figure*}
%

Whereas in viscous hydrodynamics the shear viscosity to entropy
density ratio $\eta/s$ is the important input parameter, in microscopic
transport theory the cross section or collision rate governs
the behavior of the system.
A big challenge is to connect these both parameters e.g. by
extracting the shear viscosity $\eta$ from transport calculations.
There is a big interest to calculate the shear viscosity
in a full pQCD simulation including inelastic processes,
which was done in previous calculations
\cite{Xu:2007ns,El:2008yy}. In this section we present another two
different methods to extract the shear viscosity $\eta$
numerically from transport calculations employing the
transport model BAMPS.

The first method is motivated by the classical picture of a
stationary velocity gradient. We introduce a particle system
embedded between two plates moving in opposite $z$-direction represented
by thermal reservoirs of particles with $\pm v_{\rm wall}$.
In $x$-direction the system has a size of $L$,
in $y$ and $z$ direction the system is homogeneous. We assume
the mean free path of the particles to be small compared to the
system size, i.e. $\lambda_{\rm mfp} <<L$. After a sufficient
long time scale a stationary velocity gradient $v_{\rm z}(x)$
is established.

In the non-relativistic limit the velocity gradient is linear,
$dv_{\rm z}(x)/dx = \rm const$.
Within the Navier-Stokes Ansatz the shear stress tensor in $x$-$z$-
direction is proportional to the gradient of the velocity:
\begin{equation*}
 \pi^{xz} = - \eta \frac{\partial v_z(x)}{\partial x}
\end{equation*}
In the true relativistic limit the velocity gradient is not
linear, but has the shape of a hyperbolic tangent
\begin{equation*}
v(x) = \tanh(\theta x)
\end{equation*}
 where the rapidity $y_{\rm z}(x) = \theta x$
is the argument of the hyperbolic tangent and $\theta = dy_{\rm z}/dx$.

We employ BAMPS to extract the velocity profile with
$v_{\rm wall}=0.964$. We use a massless Boltzmann gas, where the
interactions between particles are realized only via binary
collisions and isotropic cross sections. The results are shown in
Fig.\ref{fig:velocityGradient}.
On the left panel we show the velocity profile, which has the expected shape
of a hyperbolic tangent. The results are compared to
a linear fit with $\tanh(0.4x)$ according to the expected rapidity
profile shown on the right panel of Fig.\ref{fig:velocityGradient}.
For small velocities the system achieves the correct non-relativistic
limit, where the velocity gradient is linear.

Extracting the gradient of the rapidity profile $\theta$ allows us to calculate
the shear viscosity coefficient using the common relativistic relation for the
shear stress tensor
\begin{equation*}
\label{eq:shearRelativistic}
 \pi^{xz} = - \eta \gamma \theta \, ,
\end{equation*}
where $\pi^{xz}$ for a kinetic cascade can be extracted numerically (right hand side)
from BAMPS
\begin{equation*}
 \pi^{xz}({\bf r},t) = \int d^4p ~ \frac{p^1 p^3}{p^0} f({\bf p},
{\bf r} ) \cong \sum_{j=1}^N \frac{p_j^1 p_j^3}{p_j^0} \delta ({\bf
r_j} - {\bf r})
\end{equation*}
To verify the accuracy of this method the results were compared
to the analytical expression \cite{deGroot} for binary collisions and isotropic
cross sections $\eta = 1.2654 \cdot T/\sigma$, where $T$ is the temperature
and $\sigma$ is the implemented cross section. We found an
excellent agreement.


%
\begin{table*}[t!]
\begin{center}
  \begin{tabular}{| c || c | c | c | c |}
    \hline
    collision type & ($\eta/s$) shear flow & ($\eta/s$) green kubo & ($\eta/s$) Xu et al. \cite{Xu:2007ns} & ($\eta/s$) El et al. \cite{El:2008yy}\\ \hline
    $2 \leftrightarrow 2$ & - & 1.035 $\pm$ 0.015 & 1.03  & 1.4 \\ \hline
    $2 \leftrightarrow 2$ and $2 \leftrightarrow 3$  & 0.12 $\pm$ 0.07  & 0.13 $\pm$ 0.02 & 0.13  & 0.16 \\ \hline
  \end{tabular}
\caption{$\eta/s$ values extracted from BAMPS using the shear flow and Green-Kubo relation discussed in this proceeding.
Full pQCD simulations with and without inelastic $2 \leftrightarrow 3$ Bremsstrahlung processes were performed with
$\alpha_s = 0.3$ and $T= 0.4$ GeV. The extracted values are compared to the results of previous calculations
discussed in \cite{Xu:2007ns,El:2008yy}.}
\label{tb:table}
\end{center}
\end{table*}

The use of Green-Kubo relations is another possibility to extract the
shear viscosity from numerical simulations.
Green-Kubo relations connect a linear transport coefficient with the
integrated correlation function of the underlying flow connected with
the transport coefficient.
In hydrodynamics the shear stress tensor is generated by a shear
gradient. Using the Green-Kubo framework the
shear viscosity $\eta$ is found to be linear with the
correlation function of the equilibrium shear tensor.
The presence of fluctuations is attested to every equilibrated system
and Onsager's regression hypothesis \cite{PhysRev.37.405} tells that
fluctuations are driven back to equilibrium by the same transport
parameter as small non-equilibrium deviations, which motivates the
Green-Kubo relations.
The type of interaction and its property, for example the collision
rate, directly influence the emergence and decline of fluctuations.
Therefore these effects also influence the shape of the correlation
function which is the link to the shear viscosity.
With a semi-analytic discussion of the correlation function the shear
viscosity of the system can be reduced to be only dependent on the
mediums relaxation time of its auto correlation function. As a
consequence, this relaxation time includes all interaction and medium
effects.

In relativistic notation the Green-Kubo relation for the shear viscosity has the
following form \cite{roekpe}:
\begin{eqnarray*}
 \eta &&= \lim_{T \to \infty} \frac{1}{2 T} \cdot \frac{V}{10 \cdot k_B T} 
\int_{-\infty}^{\infty} d \tau \int_V d{\bf r}\int_V d{\vec x} \\
&& \times  \int_{-T}^{T} dt \left \langle \pi^{xz}(\vec x,t) \cdot \pi^{xz}(\vec x
+ \vec r ,t+\tau) \right \rangle\, ,
\end{eqnarray*}
where $\pi^{xz}({\bf r},t)$ is defined above.

The expression for the shear viscosity can be brought in a more convenient
form when the correlation function of shear tensor is known.
BAMPS solves the Boltzmann equation for an ultra relativistic gluon gas with
stochastically interpreted cross sections. The observed correlation function
follows an exponential decay $Cor(\tau) = C_0 \cdot \exp{-t / \tau}$. The zero
time correlation $C_0$ can be derived analytically by the variance of the
equilibrium Boltzmann distribution, the equilibrium gluon density and its
projection on the xy-plane:
\begin{equation*}
 Cor(\tau) = \int_V Cor({\bf r}, \tau=0) d{\bf r} = \left ( \frac{4}{\pi}
\right ) T^5 \cdot \frac{1}{V} \cdot \exp{-t / \tau}
\end{equation*}
With the exponential ansatz and the zero time correlation, the Green Kubo
relation can be integrated analytically.
To calculate the shear viscosity to entropy density, the equilibrium
density for an ultra relativistic gas is used: $s = 4n$.
This will give the simplified Green-Kubo relation for BAMPS, which
depends only on $\tau$:
\begin{equation*}
  \frac{\eta}{s} = \left ( \frac{\pi}{20} \right ) \cdot T \cdot \tau
\end{equation*}
$\tau$ includes all interaction effects of the medium, see
Fig. \ref{fig:iso_cs22}
for a correlation functions with isotropic and constant cross sections.
Again, this method is tested for binary and isotropic cross sections
and comparison gives an excellent agreement to the known results.


Finally, in Table \ref{tb:table} we now summarize the results extracted
from full pQCD simulations performed with BAMPS using both introduced methods.
We use a coupling constant $\alpha_s = 0.3$ and $T= 0.4$ GeV and show
the results with and without inelastic pQCD Bremsstrahlung. In addition
the results are compared to previous calculations, based on
Navier Stokes \cite{Xu:2007ns} and Grads method
\cite{El:2008yy}. The results presented in \cite{El:2008yy}
are extracted from full dynamical simulations, therefore loss of chemical equilibrium
and the size of equilibrium deviations might have an effect on the
obtained values. In contrast, the methods discussed in these proceedings
employ a static setup. We observe an overall good agreement.
For $\alpha_s = 0.3$ the shear viscosity to entropy density ratio is
around $\eta/s = 0.13$ as raised in the introduction and in \cite{Xu:2007ns};
this low value is mainly due to the implementation of inelastic
$2 \leftrightarrow 3$ Bremsstrahlung processes \cite{Xu:2007ns},
which are thus of crucial relevance to understand the perfect liquid
behavior of the system.

A future challenge will be to extract the shear viscosity $\eta$ for much small
coupling constant $\alpha_s$ and to check \cite{wespReining} for the
expected $\eta \sim \alpha_s^2 \log(1/\alpha_s)$ behavior \cite{PhysRevLett.64.1867}.

%

\end{document}